\documentclass[aps,pre,showpacs,amsmath,amssymb,amsfonts,superscriptaddress,lengthcheck,twocolumn,longbibliography]{revtex4-2}

\usepackage{color}
\usepackage{graphicx}
\usepackage{amsthm}
\usepackage{subfigure}
\usepackage{tabu}
\usepackage{dcolumn}
\usepackage{bm}
\usepackage{epsf}
\usepackage{color}
\usepackage[colorlinks=true,citecolor=blue,linkcolor=blue,urlcolor=blue]{hyperref}
\usepackage{hhline}
\usepackage{mathtools}
\usepackage{bbold}

\newcommand{\be}{\begin{equation}}
\newcommand{\ee}{\end{equation}}
\newcommand{\ba}{\begin{align}}
\newcommand{\ea}{\end{align}}
\newcommand{\bi}{\begin{itemize}}
\newcommand{\ei}{\end{itemize}}

\newcommand{\la}{\left\langle}
\newcommand{\ra}{\right\rangle}
\newcommand{\pd}{\partial}

\newcommand{\bla}{bla\\bla\\bla\\bla\\bla}

\newcommand{\mb}[1]{\mbox{\boldmath$#1$}}
\newcommand{\mc}[1]{\mathcal{#1}}

\DeclarePairedDelimiterX{\infdivx}[2]{(}{)}{%
  #1\;\delimsize\|\;#2%
}

\begin{document}

\title{Optimal finite-time processes in weakly driven overdamped Brownian motion}

\author{Pierre Naz\'e}
\email[]{p.naze@ifi.unicamp.br}
\affiliation{Instituto de F\'isica `Gleb Wataghin', Universidade Estadual de Campinas, 13083-859, Campinas, S\~{a}o Paulo, Brazil}

\author{Sebastian Deffner}
\email[]{deffner@umbc.edu}
\affiliation{Department of Physics, University of Maryland, Baltimore County, Baltimore, MD 21250, USA}
\affiliation{Instituto de F\'isica `Gleb Wataghin', Universidade Estadual de Campinas, 13083-859, Campinas, S\~{a}o Paulo, Brazil}

\author{Marcus V. S. Bonan\c{c}a}
\email[]{mbonanca@ifi.unicamp.br}
\affiliation{Instituto de F\'isica `Gleb Wataghin', Universidade Estadual de Campinas, 13083-859, Campinas, S\~{a}o Paulo, Brazil}

\date{\today}

\begin{abstract}

The complete physical understanding of the optimization of the thermodynamic work still is an important open problem in stochastic thermodynamics. We address this issue using the Hamiltonian approach of linear response theory in finite time and weak processes. We derive the Euler-Lagrange equation associated and discuss its main features, illustrating them using the paradigmatic example of driven Brownian motion in overdamped regime. We show that the optimal protocols obtained either coincide, in the appropriate limit, with the exact solutions by stochastic thermodynamics or can be even identical to them, presenting the well-known jumps. However, our approach reveals that jumps at the extremities of the process are a good optimization strategy in the regime of fast but weak processes for any driven system. Additionally, we show that fast-but-weak optimal protocols are time-reversal symmetric, a property that has until now remained hidden in the exact solutions far from equilibrium. 


\end{abstract}

\maketitle

\section{Introduction}
\label{sec:intro}

The great development of experimental techniques in the last decades has increased our control of systems formed of few atoms or molecules \cite{chu2002control,vander2005control,koch2019control,kumar2020nature,deffner2020natphys}. From optical lattices to superconducting devices, the quest of efficient control, in the sense of reducing general costs, has become an actual problem with several applications. However, realistic processes occur in finite time, and hence drive the system of interest out of equilibrium, implying an unavoidably higher cost than their quasistatic counterparts. The problem of finding the finite-time control with the minimum possible energetic cost, is posed then as a current challenge \cite{deffner2020}. 

To find the optimal finite-time processes is in general a very hard task and only few examples have exact solutions for arbitrary regimes. One of these is the paradigmatic case of driven Brownian motion \cite{seifert2007,seifert2008}. The existing experimental implementations of this system certify its relevance for non-equilibrium phenomena. Using colloidal particles trapped by optical tweezers, driven Brownian motion has been used to address fluctuation theorems \cite{blickle2006prl,ciliberto2008epl,ciliberto2008jstatmech}, heat engines \cite{blickle2012nature,proesmans2016prx,roldan2016soft}, feedback processes \cite{toyabe2010experimental}, Maxwell's demons \cite{roldan2014natphys}, and bit erasure \cite{berut2012nature,jun2014prl,gavrilov2016prl}.  

Despite the existence of exact optimal finite-time processes in driven Brownian motion \cite{seifert2007,seifert2008}, the physics of these processes is not very well understood. The optimal protocols present unexpected features such as jumps and sharp peaks that have been barely understood physically and represent a real challenge for experimental implementation. Such counter-intuitive characteristics have been reproduced by numerical approaches based on optimal control \cite{engel2008,geiger2010pre,aurell2011prl} but this approach has not clarified the role of such features in the optimization of the energetic cost. In other words, it is not clear how these features help to decrease the energy spent and whether they should also work as a good strategy in other driven systems.

On the other hand, perturbative approaches have been developed to provide approximate optimal finite-time protocols. Such approaches contrast with most of the numerical methods directly applied to optimize the very first definition of the energetic cost. Instead, they try to express such a cost as a functional of the corresponding finite-time protocol in terms of quantities that might describe the causes of dissipation. Although restricted to limited non-equilibrium regimes, these formulations might provide a better physical intuition about the optimal processes. Among these perturbative formulations, the so-called geometric one has attracted considerable attention in the last decade \cite{crooks2012,deweese2012,deffner2014optimal,zulkowski2015pre,zulkowski2015pre2,sivak2016pre,rotskoff2016pre,rotskoff2017pre,lucero2019pre,scandi2019quantum,blaber2020jcp,llobet2020,louwerse2022preprint,blaber2022preprint,frim2022,abiuso2022,wadiaPRE2022}. It has been applied to different non-equilibrium situations in biophysics \cite{lucero2019pre,blaber2020jcp,blaber2022preprint}, magnetic systems \cite{rotskoff2016pre,rotskoff2017pre,louwerse2022preprint}, heat engines \cite{llobet2020}, and solid-state physics \cite{antonelli1997}, in addition to having been extended to quantum systems \cite{zulkowski2015pre2,scandi2019quantum}. In this approach, the energetic cost is written as the time integral of a Lagrangian, which can be understood as a thermodynamic metric \cite{salamon1983prl,ruppeiner1995rmp,crooks2007prl}. The optimal finite-time processes are then the corresponding geodesics. 

In this work, we explore the results of another perturbative approach based on linear response theory \cite{bonanca2015,deffner2018}. It describes arbitrarily fast but necessarily weak processes. By weak we mean that the difference between the final and initial values of the control parameter must be small when compared to the initial value. What we show here contrasts however with part of the literature in stochastic thermodynamics where linear response is understood as the use of Onsager-Casimir relations \cite{brandner2015prx,brandner2016pre}. For fast and strong processes in driven Brownian motion, it has been shown that the optimal protocols obtained from the approach considered here perform better than those obtained from the geometric one \cite{kamizaki2022}. One of the main results we provide is the fact that optimal protocols must be time-reversal symmetric in the regime of weak driving. In other words, these protocols must be identical to their time-reversed counterparts. This property has been scarcely mentioned in the the literature although it can be noticed also far from equilibrium by inspecting the exact results derived in Refs.~\cite{seifert2007,seifert2008}.


\section{Finite-time optimal protocols}
\label{sec:ftopt}

\subsection{Linear response theory}
\label{subsec:lrt}

Consider a classical system, described by a Hamiltonian $\mc{H}(\mb{z_0},\lambda(t))$ and formed by a system of interest and its heat bath. The symbols $\mb{z_0}$ and $\lambda(t)$ denote respectively a point in the phase space $\Gamma$ of the whole system and a time-dependent external parameter. Initially, the system is at temperature $\beta\equiv {(k_B T)}^{-1}$, where $k_B$ is Boltzmann's constant. During a switching time $\tau$, the external parameter is changed from $\lambda_0$ to $\lambda_0+\delta\lambda$. The average work performed by the system during this process is \cite{jarzynski2007crp}
\be
W \equiv \int_0^\tau \overline{\pd_{\lambda}\mc{H}}(t)\dot{\lambda}(t)dt,
\label{eq:work}
\ee
where $\partial_\lambda$ is the partial derivative in respect to $\lambda$ and the superscript dot denotes the total time derivative. The generalized force $\overline{\pd_{\lambda}\mc{H}}$ is calculated using the average over the non-equilibrium distribution of the whole system $\rho(\mb{z_0},t)$,
\be
\overline{\partial_{\lambda}H}(t) = \int_{\Gamma}\partial_{\lambda}H(\mb{z_0})\rho(\mb{z_0},t)d{\mb{z_0}}\,.
\ee
This non-equilibrium distribution $\rho(\mb{z_0},t)$ evolves according to Liouville's equation
\be
\dot{\rho} = -\{\rho,\mc{H}\},
\ee
where $\{\cdot,\cdot\}$ is the Poisson bracket. As mentioned before, the initial distribution $\rho(\mb{z_0},0)$ is assumed to be a canonical ensemble. Consider also that the time variation of the  external parameter can be expressed as
\be
\lambda(t) = \lambda_0+g(t)\delta\lambda,
\ee
where the protocol $g(t)$ must satisfy the following boundary conditions
\be
g(0)=0,\quad g(\tau)=1\,. 
\label{eq:bc}
\ee
since $\lambda$ is switched from $\lambda_{0}$ to $\lambda_{0}+\delta\lambda$. Also, we consider that $g(t)\equiv g(t/\tau)$, which means that the intervals of time are measured in units of the switching time $\tau$.

Linear response theory aims to express non-equilibrium averages up to the first order in some perturbation parameter considering how the perturbation affects the observable to be averaged and the non-equilibrium ensemble \cite{kubo1985}. In our case, we consider that the parameter does not change significantly during the process, i.e., $|g(t)\delta\lambda/\lambda_0|\ll 1$, for all $t\in[0,\tau]$. Using the framework of linear response theory, the generalized force can be approximated up to first order as \cite{bonanca2015}
\begin{equation}
\begin{split}
\overline{\pd_{\lambda}\mc{H}}(t) =&\, \la\pd_{\lambda}\mc{H}\ra_0+\delta\lambda\la\pd_{\lambda\lambda}^2\mc{H}\ra_0 g(t)\\
&-\delta\lambda\int_0^t \phi_0(t-t')g(t')dt',
\label{eq:genforce-resp}
\end{split}
\end{equation}
where $\la\cdot\ra_0$ denotes the average over the initial canonical ensemble $e^{-\beta\mathcal{H}(\lambda_{0})}/\mathcal{Z}(\lambda_{0})$, where $\mathcal{Z}(\lambda_{0})$ is the partition function. The quantity $\phi_0(t)$ is the so-called response function \cite{kubo1985}, which can be conveniently expressed as the derivative of the relaxation function $\Psi_0(t)$ \cite{kubo1985}
\be
\phi_0(t) = -\frac{d \Psi_0}{dt}.
\ee 
In our particular case, the relaxation function reads \cite{kubo1985}
\be
\Psi_0	(t) = \beta\la\pd_\lambda\mc{H}(0)\pd_\lambda\mc{H}(t)\ra_0-\mc{C},
\label{eq:relaxfuncdefauto}
\ee 
where the constant $\mc{C}$ guarantees that $\Psi_{0}(t)$ vanishes in the limit $t\to\infty$ (we assume here that the coupling to the heat bath ensures the decay of the auto-correlation function in Eq. \eqref{eq:relaxfuncdefauto}) \cite{kubo1985, bonanca2015}. We define the relaxation time $\tau_R$ of the system as
\be
\tau_R(\lambda_{0}) = \int_0^\infty\frac{\Psi_0(t)}{\Psi_0(0)}dt.
\label{eq:relaxtimedef}
\ee

The generalized force, written in terms of the relaxation function, reads
\begin{equation}
\begin{split}
\overline{\pd_{\lambda}\mc{H}}(t) =&\, \la\pd_{\lambda}\mc{H}\ra_0-\delta\lambda\widetilde{\Psi}_0 g(t)\\
&+\delta\lambda\int_0^t \Psi_0(t-t')\dot{g}(t')dt',
\label{eq:genforce-relax}
\end{split}
\end{equation}
where $\widetilde{\Psi}_0(t)\equiv \Psi_0(0)-\la\pd_{\lambda\lambda}^2\mc{H}\ra_0$. Finally, combining Eqs. (\ref{eq:work}) and (\ref{eq:genforce-relax}), the average work performed up to linear order of the generalized force is
\begin{equation}
\begin{split}
W = &\, \delta\lambda\la\pd_{\lambda}\mc{H}\ra_0-\frac{\delta\lambda^2}{2}\widetilde{\Psi}_0\\
&+\delta\lambda^2 \int_0^\tau\int_0^t \Psi_0(t-t')\dot{g}(t')\dot{g}(t)dt'dt.
\label{eq:work2}
\end{split}
\end{equation}

It can be shown that the double integral in Eq. (\ref{eq:work2}) vanishes for $\tau\rightarrow\infty$ \cite{naze2020}. Hence, the first two terms on the right hand side must correspond to the free-energy difference (for small $\delta\lambda/\lambda_{0}$) between the final and initial equilibrium states, since this quantity is exactly the average work performed for quasistatic processes (See App.~\ref{app:A} for a detailed discussion). Thus, Eq. \eqref{eq:work2} splits into two contributions, namely,
\be
\Delta F = \delta\lambda\la\pd_{\lambda}\mc{H}\ra_0-\frac{\delta\lambda^2}{2}\widetilde{\Psi}_0,
\ee  
and
\begin{equation}
\begin{split}
W_{\text{irr}} = \delta\lambda^2 \int_0^\tau\int_0^t \Psi_0(t-t')\dot{g}(t')\dot{g}(t)dt'dt.
\label{eq:wirrder0}
\end{split}
\end{equation}
In particular, the irreversible work $W_{\rm irr}$ can be rewritten using the parity of the relaxation function, $\Psi_{0}(-t)=\Psi_{0}(t)$ \cite{kubo1985},
\begin{equation}
\begin{split}
W_{\text{irr}} = \frac{\delta\lambda^2}{2} \int_0^\tau\int_0^\tau \Psi_0(t-t')\dot{g}(t')\dot{g}(t)dt'dt.
\label{eq:wirrder}
\end{split}
\end{equation}
We aim to find the optimal protocol $\lambda(t)$ that optimizes the $W_{\rm irr}$ given by Eq. \eqref{eq:wirrder}. Hence, from now on, we focus on the minimization of this functional.

\begin{figure}
    \includegraphics[scale=0.45]{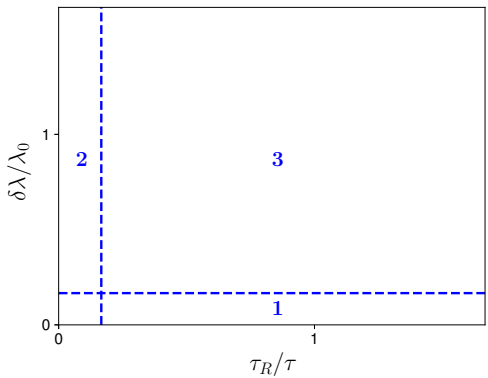}
    \caption{(Color online) Diagram of nonequilibrium regions. In region 1, we have finite-time but weak processes. In region 2, the slowly-varying processes and region 3 contains arbitrary far-from-equilibirum processes. The driving is called sudden when $\tau_R/\tau \rightarrow\infty$.}
\label{fig:diagram}
\end{figure}

Before discussing the Euler-Lagrange equation that furnishes the optimal protocol, we want to briefly emphasize what is the out-of-equilibrium regime we are describing. Near-equilibrium regimes are determined by the relative strength of the driving in respect to the initial value of the protocol, $\delta\lambda/\lambda_0$, and the rate by which the process occurs in respect to the relaxation time $\tau_R$ of the system, i.e., $\tau_R/\tau$. See Fig. \ref{fig:diagram} for a diagram depicting the regimes. In region 2, where we have the so-called slowly-varying processes, the ratio $\delta\lambda/\lambda_0$ is arbitrary, while $\tau_R/\tau\ll 1$. By contrast, in region 1, where we have finite-time and weak processes, $\delta\lambda/\lambda_0\ll 1$ while $\tau_R/\tau$ is arbitrary. In region 3, we have far-from-equilibrium processes and both ratios are arbitrary. In particular, in region 1, drivings where the ratio $\tau_R/\tau\rightarrow\infty$ are called sudden processes. In the present work, we will focus on processes lying in region 1. See Refs. \cite{mou1994,antonelli1997,crooks2012,deweese2012,deffner2014optimal} for more details on slowly-varying processes and App.~\ref{app:B} to see how both approaches, from region 1 and 2, converge when the appropriate limits are taken. 

\subsection{Euler-Lagrange equation}
\label{subsec:ele}

The functional \eqref{eq:wirrder} is not written in the most convenient way since the specified boundary conditions, $g(0)=0$ and $g(\tau)=1$, do not apply to $\dot{g}(t)$. To circumvent this problem, we rewrite the irreversible work in terms of $g(t)$. Using integration by parts in Eq. (\ref{eq:wirrder}), we have
\begin{equation}
\begin{split}
W_{\text{irr}} &= \frac{\delta\lambda^2}{2}\Psi_0(0)+\delta\lambda^2\int_0^\tau\dot{\Psi}_0(\tau-t')g(t')dt'\\
&-\frac{\delta\lambda^2}{2}\int_0^\tau \int_0^\tau\ddot{\Psi}_0(t-t')g(t)g(t')dtdt',
\label{eq:wirr}
\end{split}
\end{equation}
where we have used the boundary conditions \eqref{eq:bc} and a direct consequence of the parity property of the relaxation function, $\dot{\Psi}(0)=0$. Using calculus of variations \cite{gelfand2000,kirk2004,liberzon2011}, the Euler-Lagrange equation is
\begin{equation}
\int_0^\tau \ddot{\Psi}_0(t-t')g^*(t')dt' = \dot{\Psi}_0(\tau-t),
\label{eq:eleq}
\end{equation}
where the optimal protocol $g^{*}(t)$ must satisfy $g^{*}(0)=0$ and $g^{*}(\tau)=1$. It is remarkable that the optimal protocol depends only on characteristics of the system in the initial equilibrium state. For instance, the previous equation does not depend on the driving strength $\delta\lambda/\lambda_0$.

Some other aspects however are not completely clear: since solutions with jumps exist for thermodynamics processes \cite{seifert2007,engel2008}, how exactly can such solutions be obtained from Eq. (\ref{eq:eleq})? How exactly do they satisfy the boundary conditions of Eq. (\ref{eq:bc})? Are these jumps included in the functional of Eq. (\ref{eq:wirr})? In Sec.~\ref{subsec:jumps}, we show how Eq. \eqref{eq:eleq} can admit  optimal protocols having jumps at their extremities. 


As a last remark of this section, we show that simple functions satisfy Eq. \eqref{eq:eleq} in the limits of extremely long and extremely short protocols. Consider, for instance, the linear protocol $g(t)=t/\tau$. Substituting it in the left hand side of Eq. \eqref{eq:eleq} and performing integration by parts, we obtain
\be
\int_0^\tau \ddot{\Psi}_0(t-t')\frac{t'}{\tau}dt' = \dot{\Psi}_0(\tau-t)+\frac{1}{\tau}(\Psi_0(\tau-t)+\Psi_0(t)),
\label{eq:elslowly}
\ee
which satisfies the Euler-Lagrange equation when $\tau\rightarrow\infty$, since the last two terms in the right hand side go to zero in this limit. For sudden processes, where $\tau\rightarrow 0$, the Euler-Lagrange equation is satisfied for a constant protocol $g^{*}(t)=1/2$
\be
\frac{1}{2}\int_0^\tau \ddot{\Psi}_0(t-t') dt' = \frac{1}{2}(\dot{\Psi}_0(\tau-t)+\dot{\Psi}_0(t)). 
\ee

Mathematically, Eq. (\ref{eq:eleq}) is an inhomogeneous Fredholm equation of the first kind \cite{courant2008}. This kind of equation describes what is called {\it ill-posed} problems, where the existence, uniqueness, and stability of solutions are not completely guaranteed \cite{groetsch2007}. Nevertheless, from a practical point-of-view, solutions for particular kernels can be found in handbooks of integral equations \cite{polyanin2008}.

\subsection{Contributions of jumps to the irreversible work}
\label{subsec:jumps}

\begin{figure}
    \includegraphics[scale=0.45]{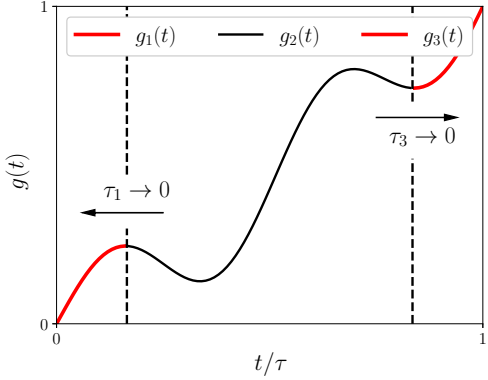}
    \caption{(Color online) Schematic representation of the protocol $g(t)$ in three parts. The red curves at the initial and final parts become sudden processes when $\tau_{1}$ and $\tau_{2}$ are taken close to zero. In this manner, the boundary conditions $g(0)=0$ and $g(\tau)=1$ always hold.}
\label{fig:setup}
\end{figure}

To evaluate how jumps affect the optimization of $W_{\rm irr}$, we will reformulate the previous analysis including explicitly these features in the protocol. Consider a protocol $g(t)$, with $t\in[0,\tau_1+\tau_2+\tau_3]$, composed of three parts denoted by $g_{i}$, with $i=1,2,3$, each one with duration $\tau_{i}$, and starting at different inital times $\Delta\tau_{i-1}$,
\be
h_i(t)\equiv g_i\left(\left(t-\Delta\tau_{i-1}\right)/\tau_i\right),\quad t\in\left[\Delta\tau_{i-1},\Delta\tau_{i}\right],
\ee
where 
\be
\Delta\tau_i = \sum_{k \le i}\tau_k,
\ee
and we take $\Delta\tau_{0}=0$. The following boundary and continuity conditions must be satisfied
\be
g_1(0)=0,\quad g_3(1)=1,
\ee
\be
g_1(1)=g_2(0),\quad g_2(1)=g_3(0),
\label{eq:bc2}
\ee
where Eqs. (\ref{eq:bc2}) guarantee that the whole process is continuous although not necessarily smooth at all points. In particular, we consider that $\tau_1\rightarrow 0$ and $\tau_3\rightarrow 0$, so that protocols $g_1$ and $g_3$ become sudden processes. Figure \ref{fig:setup} depicts what is meant by this. 

The irreversible work of a process consisting of three pieces can be written as follows
\be
W_{\text{irr}} = \sum_{i=1}^{3}\sum_{j\le i}W_{\text{irr}}^{ij},
\label{eq:wirrtotal}
\ee
where
\be
W_{\text{irr}}^{ij} = \delta\lambda^2\int_{\Delta\tau_{i-1}}^{\Delta\tau_{i}} \dot{h}_i(t)f_j(t)dt,
\ee
and
\be
f_j(t) = 
\begin{cases}
\displaystyle\int_{\Delta\tau_{j-1}}^{\Delta\tau_{j}} \Psi_0(t-t')\dot{h_j}(t')dt',\quad j<i,\\
\displaystyle\int_{\Delta\tau_{j-1}}^{t} \Psi_0(t-t')\dot{h_j}(t')dt',\quad j=i.\\
\end{cases}
\label{eq:wirrparts}
\ee
In Appendix \ref{app:C}, we show in details how Eqs. \eqref{eq:wirrtotal} to \eqref{eq:wirrparts} are obtained. After the limits $\tau_{1,3}\to 0$ are taken, the irreversible work reads
\begin{equation}
\begin{split}
W_{\text{irr}} & = \frac{\delta\lambda^2}{2}\Psi_0(0)+\delta\lambda^2\int_0^{\tau_2}\dot{\Psi}_0(\tau-t')g_2(t'/\tau_2)dt'\\
&-\frac{\delta\lambda^2}{2}\int_0^{\tau_2} \int_0^{\tau_2}\ddot{\Psi}_0(t-t')g_2(t/\tau_2)g_2(t'/\tau_2)dtdt',
\label{eq:wirrtotal2}
\end{split}
\end{equation}
which is identical to the functional of Eq. (\ref{eq:wirr}). This means that the boundary conditions \eqref{eq:bc} do not have to be imposed to the solutions of the Euler-Lagrange equation \eqref{eq:eleq}. 

In other words, the solutions of Eq. \eqref{eq:eleq}, with free boundary conditions, furnish the optimal protocols and their corresponding jumps. Thus, we are able to show that small jumps at the beginning and at the end of the protocol are costless and become an unexpected strategy to reduce $W_{\rm irr}$. So far, this had been shown exactly only for driven Brownian motion in Refs.~\cite{seifert2007,seifert2008}. Here we have shown that this is true for any driven system in the regime of fast but weak processes.

\subsection{Time reversal symmetry}
\label{subsec:time}

Reference~\cite{deffner2018} suggested that fast-but-weak optimal protocols are time-reversal symmetric. In this subsection, we will prove this statement. We first comment what  we mean by time-reversal symmetry. Consider a forward process, denoted by $\lambda_F(t)$, that takes the system from $\lambda_0$ to $\lambda_0+\delta\lambda$, according to the forward protocol $g_F(t)$. The time-reversed process associated to $\lambda_{F}(t)$, denoted by $\lambda_R(t)$, is defined as $\lambda_R(t)\equiv \lambda_F(\tau-t)$, which takes the system from $\lambda_0+\delta\lambda$ to $\lambda_0$. In App.~\ref{app:D}, we show that the time-reversed protocol $g_{R}(t)$ is given by
\be
g_R(t) = 1-g_F(\tau-t).
\label{eq:backforw}
\ee
A protocol that presents what we call time-reversal symmetry if it is identical to its time-reversed twin, that is, $g_F(t)=g_R(t)$.

We will show now that the solutions $g^{*}(t)$ of Eq. \eqref{eq:eleq} are time-reversal symmetric. To do that, we explicitly show that the time reversal of $g^*(t)$ is also a solution of Eq.~\eqref{eq:eleq}. Plugging $1-g^*(\tau-t)$ into Eq. \eqref{eq:eleq} we obtain
\be
\int_0^\tau \ddot{\Psi}_0(t-t')[1-g^*(\tau-t')]dt' = \dot{\Psi}_0(\tau-t).
\label{eq:reverseg}
\ee
After some algebra, we arrive at
\be
\int_0^\tau \ddot{\Psi}_0(t-\tau+t')g^{*}(t')dt' = \dot{\Psi}_0(t).
\label{eq:reverseg2}
\ee
Using now the parity of the second derivative of the relaxation function ($\Psi_{0}(-t) = \Psi_{0}(t)$ implies $\ddot{\Psi}_{0}(-t)=\ddot{\Psi}_{0}(t)$), we arrive at
\be
\int_0^\tau \ddot{\Psi}_0(\tau-t-t')g^{*}(t')dt' = \dot{\Psi}_0(t).
\label{eq:reverseg3}
\ee
Changing the variable $t'\rightarrow \tau-t'$, we finally have
\be
\int_0^\tau \ddot{\Psi}_0(t-t')g^{*}(t')dt' = \dot{\Psi}_0(\tau-t),
\label{eq:reverseg4}
\ee
which is Eq.\eqref{eq:eleq}. Since $g^{*}(t)$ was assumed to be a solution of \eqref{eq:eleq}, this shows that $1-g^{*}(\tau-t)$ is also an optimal protocol.

This symmetry of the optimal protocol, at first glance restricted to the regime of weak processes, has been mostly neglected and reveals an important element of the physics of optimal finite-time processes. For instance, it seems to be absent in the regime described by the geometric approach as the corresponding Euler-Lagrange equation does not contain it \cite{crooks2012,deffner2014optimal}. In Sec.~\ref{sec:compstochover}, we will exemplify it using driven Brownian motion and show that it is also present in arbitrarily far-from-equilibrium processes. In the context of information erasure, the role of time-reversal symmetry has been shown relevant \cite{wimsatt2021refining}. 

We also remark that the solution $g^{*}(t)$, obtained using standard calculus of variations, is at most a local minimum. Thus, it is not ruled out the possibility that its time reversal would be associated with a different minimum than the optimal protocol $g^{*}(t)$. Hence, if we assume that the solution of the Euler-Lagrange equation (\ref{eq:eleq}) is unique, as has been suggested by numerical experiments \cite{deffner2018}, we conclude that the optimal protocols present time-reversal symmetry.

\subsection{Optimal irreversible work}
\label{subsec:optirrwork}

Substituting Eq. (\ref{eq:eleq}) into Eq. (\ref{eq:wirr}), and using integration by parts to remove the derivative of the relaxation function, the optimal irreversible work is
\be
W_{\text{irr}}^{*} = \frac{\delta\lambda^2}{2}\Psi_{0}(0)+ \frac{\delta\lambda^2}{2}\int_0^\tau\dot{\Psi}_0(\tau-t')g^*(t')dt'.
\label{eq:optwirr}
\ee
The irreversible work $\widetilde{W}^*_{\text{irr}}$ for the time reversal of $g^*(t)$, which is also an optimal protocol, is identical to the one of the optimal protocol, since the solution is unique. Indeed, is not hard to see that, for $g_{R}^{*}(t)=1-g^{*}(\tau-t)$, the irreversible work reads
\textcolor{red}{\begin{equation}
\widetilde{W}^*_{\text{irr}}=\frac{\delta\lambda^2}{2}\Psi_{0}(\tau)-\frac{\delta\lambda^2}{2}\int_0^\tau \dot{\Psi}_{0}(t)g^*(t)dt.
\label{eq:wirrequal}
\end{equation}}
Subtracting Eq. \eqref{eq:wirrequal} from Eq. \eqref{eq:optwirr}, and using the uniqueness of the optimal protocol, we have
\be
\widetilde{W}^*_{\text{irr}} = W^*_{\text{irr}}.
\ee

As $\tau$ becomes large compared to the relaxation time $\tau_{R}$, expression~\eqref{eq:wirrequal} for the optimal irreversible work converges to the value predicted by the geometric approach when $g^*(t)=t/\tau$ (see App. \ref{app:B} for the details). Hence, 
\be
\lim_{\frac{\tau_R}{\tau}\ll 1} W_{\text{irr}}^{*} = \delta\lambda^2\Psi_0(0)\frac{\tau_R}{\tau}\,.
\label{eq:wirrsv}
\ee
This result suggests that the solution of Eq.~\eqref{eq:eleq} tends to a linear function in the slowly-varying regime.

On the other hand, as $\tau/\tau_{R}\to 0$, $g^{*}(t)$ becomes a sudden process and $W^{*}_{\text{irr}}$ is calculated using in Eq.~\eqref{eq:optwirr} in the limit $\tau\rightarrow 0$ and $g^*(t)=1/2$. Hence
\be
\lim_{\frac{\tau_R}{\tau}\gg 1} W_{\text{irr}}^{*} = \frac{\delta\lambda^2}{2}\Psi_0(0).
\label{eq:wirrsudden}
\ee
Both limits just discussed will be important to check the agreement of our approach with exact results in the examples that follow. 

\subsection{Excess power}
In the geometric approach, it has been shown that the excess power along the optimal protocol is stationary when the system is driven by a single control parameter \cite{crooks2012}. In the present case, we will show that the same happens to finite-time and weak process if the optimal protocol is a linear function.

The optimal excess power can be obtained finding the function whose integration from $t=0$ to $t=\tau$ furnishes Eq.~\eqref{eq:optwirr}. This yields
\be
\mc{P}_{\rm ex}^*(t,\tau) = \delta\lambda^2 \dot{f}^*(t)\int_0^t \Psi_{0}(t-t')\dot{f}^*(t')dt' ,
\label{eq:excesspower}
\ee
where 
\be
f^*(t) = g^*(t)+2g^*(0^+)(\Theta(t)-\Theta(\tau-t))\,,
\label{eq:protdiv}
\ee
and $g^*(t)$ is the continuous part of the optimal protocol, i.e., the solution of Eq.~\eqref{eq:eleq}. Here, $H(t)$ is the Heaviside step function,
\be
\Theta(t) = \begin{cases}
0,\quad t<0\\
\frac{1}{2},\quad t= 0\\
1,\quad t>0
\end{cases}.
\ee 
The purpose of introducing $f^{*}(t)$ is to explicitly include the initial and final jumps. According to Eq.~\eqref{eq:protdiv}, one can easily check that $f^*(0)=0$ and $f^*(\tau)=1$. As we show next, this is necessary to give the correct physical meaning to the excess power calculations. For instance, the optimal protocol in the sudden limit reads (see Sec.~\ref{subsec:ele})
\be
\lim_{\frac{\tau}{\tau_R}\ll 1} f^*(t) =\frac{1}{2}+ \Theta(t)-\Theta(\tau-t)\,.
\ee
Thus, it becomes clear that without including the jumps explicitly, Eq.~\eqref{eq:excesspower} would furnish null excess power in this case.

In general, the optimal excess power \eqref{eq:excesspower} can be decomposed into two contributions, namely, one due to the continuous part of the protocol and another due to the jumps. Denoting them respectively by $\mathcal{P}_{\rm C}^*$ and $\mathcal{P}_{\rm J}^*$, we will have (see App.~\ref{app:E} for more details) 
\begin{eqnarray}
\lefteqn{\mathcal{P}_{\rm C}^*(t,\tau) =}\nonumber\\ &&\delta\lambda^2\frac{\dot{g}^{*}(t)}{2}\left(\int_0^\tau\Psi_0(t')\dot{g}^*(t')dt'+(\Psi_0(\tau)+\Psi_0(0))g^{*}(0^+)\right)\,,\nonumber\\
\end{eqnarray}
\be
\mathcal{P}_{\rm J}^*(t,\tau) = \delta\lambda^2 {g^{*}}^2(0^{+})\Psi_0(0)(\delta(t)+\delta(\tau-t)).
\ee

We remark that the irreversible work in the slowly-varying regime is recovered by Eq. \eqref{eq:excesspower}. In this limit, we have
\be
\lim_{\frac{\tau_R}{\tau}\ll 1}\mathcal{P}_{\rm J}^*(t,\tau)=\delta\lambda^2\frac{\tau_R\Psi(0)}{\tau^2}\,,
\ee
\be
\lim_{\frac{\tau_R}{\tau}\ll 1}\mathcal{P}_{\rm J}^*(t,\tau)=0.
\ee

The derivative of $\mathcal{P}_{\rm C}^*$ reads 
\begin{eqnarray}
\lefteqn{\dot{\mathcal{P}}_{\rm C}^* =}\nonumber\\ &&\delta\lambda^2\frac{\ddot{g}(t)}{2}\left(\int_0^\tau \Psi_0(t')\dot{g}^*(t')dt'+(\Psi_0(\tau)+\Psi_0(0))g(0^+)\right),\nonumber\\
\end{eqnarray}
which means that the continuous contribution of the optimal excess power is stationary only if $g^{*}(t)$ is a linear function.

Finally, due to the time-reversal symmetry of the optimal protocol, the excess power used in the backward process is the same as in the forward one. This can be seen by direct substitution of Eq.~\eqref{eq:backforw} in Eq.~\eqref{eq:excesspower}, leading to
\be
\mathcal{P}^*_{\rm C}(t,\tau)=\mathcal{P}^*_{\rm C}(\tau-t,\tau),\quad \mathcal{P}^*_{\rm J}(t,\tau)=\mathcal{P}^*_{\rm J}(\tau-t,\tau).
\ee
This result implies that
\be
\int_0^\tau\dot{\mathcal{P}}_{\rm C}^*(t',\tau)dt'=0,
\ee
for an arbitrary optimal protocol $g^{*}(t)$. This property seems to be similar to the stationarity property predicted by the geometric approach.

\section{Comparison with stochastic thermodynamics: overdamped Brownian motion}
\label{sec:compstochover}

In this section, we present a comparison between the optimization of the irreversible work using stochastic thermodynamics and linear response theory. We consider the paradigmatic example of a Brownian particle confined by optical tweezers in two situations: a moving laser trap and a stiffening trap, both in the overdamped regime. We show that our approach leads to the same optimal protocol calculated in Ref. \cite{seifert2007}. Theoretically, we consider a particle in contact with a heat bath of inverse temperature $\beta={(k_B T)}^{-1}$, whose motion is governed by the following Langevin equation
\be
\dot{x}(t)+\frac{1}{\gamma}\partial_x V(x(t),\lambda(t)) =\eta(t),
\label{eq:langevin}
\ee   
where $x(t)$ is the position of the particle at the instant $t$, $\gamma$ is the damping coefficient, $V$ is the time-dependent harmonic potential, $\lambda(t)$ is the control parameter and $\eta(t)$ is a Gaussian white noise characterized by
\be
\overline{\overline{\eta(t)}}=0, \quad \overline{\overline{\eta(t)\eta(t')}}=\frac{2}{\gamma\beta}\delta(t-t'),
\label{eq:bceta}
\ee
with $\overline{\overline{(...)}}$ being the stochastic average over several realizations.


\subsection{Moving laser trap}
\label{subsec:overmov}

For the moving laser trap, the time-dependent harmonic potential is given by 
\be
V(x(t),t)=\frac{\omega^2_0}{2}(x(t)-\lambda(t))^2, 
\label{eq:potential1}
\ee
where $\omega_0$ the natural frequency of the oscillation, and $\lambda$ the equilibrium position. During the driving, the equilibrium position is changed from $\lambda_0$ to $\lambda_0+\delta\lambda$. For this particular example, we will show that linear response corresponds to the exact dynamics.


\subsubsection{Connection with linear response theory}

In contrast to what was done in Ref. \cite{seifert2007}, we will express the irreversible work in terms of the control parameter $\lambda(t)$ instead of the average position $u(t) = \overline{\overline{x}}(t)$ of the particle. To do that, we first solve the Langevin equation (\ref{eq:langevin}) expressing $u(t)$ in terms of $\lambda(t)$,
\be
u(t) = \lambda(t)-\int_0^t e^{-\frac{\omega_0^2}{\gamma}(t-t')}\dot{\lambda}(t')dt',
\label{eq:uviastoch}
\ee
Due to Eq.~\eqref{eq:potential1}, the work is then given by 
\be
W=\int_{0}^{\tau} \dot{\lambda}(t) \overline{\overline{\partial_{\lambda}V}}dt =\omega_{0}^2\int_0^\tau\dot{\lambda}(t)(\lambda(t)-u(t))dt\,.
\label{eq:workmovlaser}
\ee
Using Eq.~\eqref{eq:uviastoch}, we have
\be
W = \omega_0^2\int_0^\tau\int_0^t e^{-\frac{\omega_0^2}{\gamma}(t-t')}\dot{\lambda}(t')\dot{\lambda}(t)dt'dt\,.
\ee
Since the change in the Helmholtz free energy is zero in this case, the irreversible work reads
\be
W_{\text{irr}} = \omega_0^2\int_0^\tau\int_0^t e^{-\frac{\omega_0^2}{\gamma}(t-t')}\dot{\lambda}(t')\dot{\lambda}(t)dt'dt.
\label{eq:movinglaserwirrexact}
\ee

Comparing Eqs. \eqref{eq:movinglaserwirrexact} and \eqref{eq:wirrder0}, we identify the relaxation function as
\be
\Psi_0(t) = \omega_0^2e^{-\frac{\omega_0^2}{\gamma}|t|},
\label{eq:psiover}
\ee
from which it can be shown that
\be
\Delta F = \delta\lambda\la\pd_{\lambda}V\ra_0-\frac{\delta\lambda^2}{2}\widetilde{\Psi}_0 = 0,
\ee
and hence the exact expression of the irreversible work has the form of our linear-response functional. Finally, using Eq. \eqref{eq:relaxtimedef}, the relaxation time of this system is
\be
\tau_R = \frac{\gamma}{\omega_0^2}.
\ee


\subsubsection{Prediction of stochastic thermodynamics result}

As the functional provided by linear-response theory coincides with that obtained from the exact non-equilibrium dynamics, we can use the approach developed in Sec. \ref{sec:ftopt} to calculate the optimal protocol. 

As Ref. \cite{deffner2018} indicated numerically, we consider that the optimal protocol $g^*(t)$ is a time-reversal symmetric linear function
\be
g^*(t) = s^*\left(t-\frac{\tau}{2}\right)+\frac{1}{2},
\label{eq:ansatz1}
\ee
with the optimal slope coefficient $s^*$ to be determined. Plugging Eqs. \eqref{eq:psiover} and \eqref{eq:ansatz1} in the Euler-Lagrange equation \eqref{eq:eleq}, the optimal slope coefficient becomes
\be
s^* = \frac{1}{\tau+2\tau_R}.
\ee
Thus, the optimal protocol reads 
\be
g^*(t) = \frac{t+\tau_R}{\tau+2\tau_R},
\label{eq:optg1}
\ee
which is identical to the optimal protocol obtained in Ref. \cite{seifert2007}. The corresponding optimal irreversible work is
\be
W^{*}_{\rm irr} = \frac{\omega_0^2\delta\lambda^2}{\tau/\tau_R+2},
\label{eq:wirropt1}
\ee
We remark that Eqs. \eqref{eq:optg1} and \eqref{eq:wirropt1} agree with the limits of slowly-varying and sudden processes analyzed in Sec.~\ref{subsec:optirrwork} when the asymptotic behaviors $\tau/\tau_R\gg 1$ and $\tau/\tau_R\ll 1$ are considered.

To calculate the excess power \eqref{eq:excesspower} along \eqref{eq:optg1}, we need to express the optimal protocol with the appropriate boundary conditions. In this way, we have
\be
f^*(t) = \frac{t+\tau_R}{\tau+2\tau_R}+\frac{2\tau_R}{\tau+2\tau_R}(\Theta(t)-\Theta(\tau-t)).
\ee
Calculating now the excess power along $f^{*}(t)$, we have
\be
\mathcal{P}^{*}_{\rm ex}(t) = \frac{\tau_R\omega_0^2\delta\lambda^2}{(\tau+2\tau_R)^2}+\frac{2\tau_R^2\omega_0^2\delta\lambda^2}{(\tau+2\tau_R)^2}(\delta(t)+\delta(\tau-t)).
\label{eq:movingexcpow}
\ee
Equation \eqref{eq:wirropt1} is easily recovered by Eq. \eqref{eq:movingexcpow}. As it was shown before, the excess power is stationary except at the extremities of the protocol. There, it has delta peaks that indicate the jumps performed by the optimal protocol. Despite of that, the irreversible work is finite.


\subsection{Stiffening trap}

For the stiffening trap, the time-dependent harmonic potential is given by
\be
V(x(t),t)=\frac{\lambda(t)}{2}x^2(t),
\ee
where $\lambda$ is the stiffening parameter. During the driving, the parameter is changed from $\lambda_0$ to $\lambda_0+\delta\lambda$. In what follows we show that the optimal protocol of linear response theory reproduces in zeroth-order approximation the same optimal protocol predicted by Ref. \cite{seifert2007}.


\subsubsection{Prediction of stochastic thermodynamics result}

According to Ref. \cite{seifert2007}, the optimal protocol of a Brownian particle subjected to the stiffening trap in the overdamped regime is
\be
g^*(t) = \frac{1}{\delta\lambda}\left(\frac{\lambda_0}{(c^* t+1)^2}-\frac{c^* \gamma }{c^* t+1}-\lambda_0\right),
\label{eq:optst}
\ee 
where
\be
c^* = \frac{\sqrt{\gamma ^2+2 \gamma  \lambda_0 \tau +\delta \lambda  \lambda_0 \tau ^2+\lambda_0^2 \tau ^2}-\gamma -\delta \lambda  \tau -\lambda_0 \tau}{2 \gamma  \tau +\delta \lambda  \tau^2+\lambda_0 \tau ^2}.
\ee
Applying expression \eqref{eq:relaxfuncdefauto} to the present case, we obtain
\be
\Psi_0(t) = \beta\la\overline{\overline{\partial_\lambda V}}(t)\partial_\lambda V(0)\ra_0-\mc{C},
\label{eq:relaxdef}
\ee
where the generalized force $\partial_\lambda V(x(t),t)$ is stochastically averaged along the path, and the constant $\mc{C}$ is calculated such that $\Psi_0(t)\rightarrow 0$ when $t\to\infty$. Since $\partial_{\lambda}V(t)=x^{2}(t)/2$, and $x(t)$ is a solution of Eq. \eqref{eq:langevin}, the previous average can be calculated using Eqs. \eqref{eq:bceta} and the fact that the initial distribution is an equilibrium one given by $\exp{(-\beta \lambda_{0}x^{2}/2)}/\mathcal{Z}$, where $\mathcal{Z}$ is the partition function. 

Equation \eqref{eq:relaxdef} then becomes
\be
\Psi_0(t) = (2\beta\lambda_0^2)^{-1}\exp{\left(-\frac{2\lambda_0}{\gamma}|t|\right)},
\ee
which yields the relaxation time 
\be
\tau_R=\frac{\gamma}{2\lambda_0}.
\label{eq:relaxtover1}
\ee
Therefore, the optimal protocol is identical to the one calculated in Sec. \ref{subsec:overmov} (see Eq. (\ref{eq:optg1})), 
\be
g^*(t) = \frac{t+\tau_R}{\tau+2\tau_R}.
\label{eq:optg2}
\ee
but with the relaxation time given by \eqref{eq:relaxtover1}.

\begin{figure}
	\includegraphics[scale=0.45]{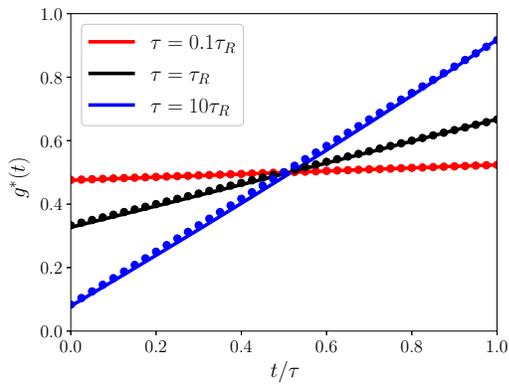}
    \caption{(Color online) Comparison between the optimal protocols for the stiffening trap in the overdamped regime. The solid lines correspond to the results predicted by stochastic thermodynamics, Eq. \eqref{eq:optst}, while the dots correspond to the linear-response result, Eq. \eqref{eq:optg2}. It was used the ratios $\tau/\tau_R = 0.1,1,10$ with $\tau_R=1/2$, considering $\gamma=\lambda_0=1$ and $\delta\lambda=0.1$.}
\label{fig:over}
\end{figure}

The optimal value of the irreversible work reads
\be
W_{\rm irr}^* = \frac{1}{2\beta}\frac{(\delta\lambda/\lambda_0)^2}{\tau/\tau_R+2}\,,
\label{eq:wirropt2}
\ee
whose asymptotic limits agree with the slowly-varying and sudden cases of Sec.~\ref{subsec:optirrwork}. One can show that Eq. (\ref{eq:optg2}) is the zeroth-order approximation of Eq. (\ref{eq:optst}) in $\delta\lambda$. Figure \ref{fig:over} illustrates this agreement for a weak perturbation of $\delta\lambda/\lambda_0=0.1$. In this case, the expression for the excess power is similar to the one presented in the moving laser trap example. 

Additionally, we remark that the exact optimal protocol \eqref{eq:optst} also presents the time-reversal symmetry discussed in Sec.~\ref{subsec:time} when it drives the Brownian particle far from equilibrium. Indeed, it can be verified, using Eq.~\eqref{eq:optst}, that
\be
g^*(t)+g^*(\tau-t) \approx 1,
\ee
when $\delta\lambda/\lambda_{0}\gg 1$ and $\tau_{R}/\tau \gtrsim 1$.

\section{Final remarks}
\label{sec:final} 

In this work, we presented an analytical approach to the optimization of the thermodynamic irreversible work for finite-time and weak processes in driven Brownian motions. We have derived the Euler-Lagrange equation that leads to the extremum of the optimization problem and have shown that it already accounts for jumps at the extremities of the protocol. As a consequence, the solutions of this equation do not require fixed boundary conditions. Thus, we have confirmed that such jumps are an unexpected strategy to reduce the irreversible work. We have also shown that the optimal protocols in this linear-response regime are time-reversal symmetric due to the properties of the relaxation function. This symmetry had been suggested before by semi-analytical results but we were able to show it in this work through the derivation of the Euler-Lagrange equation. We have also emphasized that it is present in the exact optimal protocols far from equilibrium. This shows that perturbative approaches can reveal important information about the physics of optimal processes in contrast to crude numerical methods. We have also analyzed the excess power along the optimal protocol and have found that its qualitative behavior surprisingly agrees with that observed in the slowly-varying regime. Our main results were illustrated using the paradigmatic example of driven Brownian motion. We have shown that, whenever an analytical solution from stochastic thermodynamics is available, our results coincide with them in the linear-response regime. The analysis of the underdamped regime is a natural sequel of this research.

\section*{Acknowledgements}

P.N. and M.V.S.B. acknowledge financial support from FAPESP (Funda\c{c}\~{a}o de Amparo \`a Pesquisa do Estado de S\~ao Paulo) (Brazil) (Grant No. 2018/06365-4, No. 2018/21285-7 and No. 2020/02170-4) and from CNPq (Conselho Nacional de Desenvolvimento Cient\'ifico e Pesquisa) (Brazil) (Grant No. 141018/2017-8). S. D. acknowledges support from the U. S. National Science Foundation under Grant No. DMR-2010127.


\appendix

\section{Difference of Helmholtz free energy in linear response theory}
\label{app:A}

In this appendix, we will show that what we call the free-energy difference in linear response theory,
\be
\Delta F = \delta\lambda\la\pd_{\lambda}\mc{H}\ra_0-\frac{\delta\lambda^2}{2}\widetilde{\Psi}_0,
\label{eq:deltaf2}
\ee
is indeed consistent with the definition
\be
\Delta F = \Delta U-T\Delta S
\label{eq:dfreeenergydef}
\ee
when $\delta\lambda/\lambda_{0} \ll 1$. 

The difference of the state functions in Eq.~(\ref{eq:dfreeenergydef}) are calculated between the initial and final states corresponding to $\lambda_{0}$ and $\lambda_{0}+\delta\lambda$, respectively. The term $\Delta U$ is then the difference of internal energy given by
\be
\Delta U = \int_{\Gamma}\mathcal{H}(\lambda_0+\delta\lambda)\rho_c(\lambda_0+\delta\lambda)d\Gamma-\int_{\Gamma}\mathcal{H}(\lambda_0)\rho_c(\lambda_0)d\Gamma
\label{eq:dinternalenergydef}
\ee
where $\rho_{c}(\lambda) = \exp{(-\beta \mathcal{H}(\lambda))}/\mathcal{Z}(\lambda)$ denotes the canonical distribution. In its turn, the term $\Delta S$ is the relative entropy between the final and initial equilibrium state. More precisely
\be
\Delta S =-k_B \int_\Gamma \rho_c(\lambda_0)\log{\frac{\rho_c(\lambda_0+\delta\lambda)}{\rho_c(\lambda_0)}}d\Gamma,
\label{eq:drelativeentropydef}
\ee
where $k_B$ is the Boltzmann's constant. 

Expanding Eq.~\eqref{eq:dinternalenergydef} up to second-order in $\delta\lambda$, the internal energy can be expressed as
\begin{equation}
\begin{split}
\Delta U = \delta\lambda \int_\Gamma \partial_\lambda(\mathcal{H}&(\lambda_0)\rho_c(\lambda_0))d\Gamma\\&+\frac{\delta\lambda^2}{2}\int_\Gamma \partial^2_{\lambda\lambda}(\mathcal{H}(\lambda_0)\rho_c(\lambda_0))d\Gamma\,.
\end{split}
\end{equation}
Thus,
\be
\Delta U =  \delta\lambda\la\pd_{\lambda}\mc{H}\ra_0+\frac{\delta\lambda^2}{2}\langle\partial^2_{\lambda\lambda}\mathcal{H}\rangle_0.
\label{eq:deltaufinal}
\ee

Performing the same expansion in Eq.~\eqref{eq:drelativeentropydef}, we obtain
\begin{equation}
\begin{split}
T\Delta S=-&\frac{\delta\lambda}{\beta}\left[\int\rho_{c} ^{\prime}(\lambda_0)d\Gamma+\int\rho_{c}^{\prime\prime}(\lambda_0)d\Gamma\right]
\\&+\frac{\beta\delta\lambda^2}{2}\left[\int_\Gamma \rho_c(\lambda_0)\left(\partial_{\lambda}\mathcal{H}(\lambda_0)\right)^{2}d\Gamma\right],
\end{split}
\end{equation}
where $\rho_{c}^{\prime}$ and $\rho_{c}^{\prime\prime}$ denote the first and second derivatives of the canonical distribution in respect to $\lambda_0$. Using the normalisation of $\rho_c$ and the definition of the relaxation function, we have
\be
T\Delta S = \frac{\delta\lambda^2}{2}\Psi(0).
\label{eq:deltasfinal}
\ee
Summing Eqs. \eqref{eq:deltaufinal} and \eqref{eq:deltasfinal} we arrive at Eq. \eqref{eq:deltaf2}.

\section{Compatibility between slowly-varying and finite-time and weak processes}
\label{app:B}

We denote the expressions of the for slowly-varying and finite-time and weak processes respectively as $W_{\text{irr}}^2$ and $W_{\text{irr}}^1$. The first one can be expressed as \cite{crooks2012,deffner2014optimal}
\be
W_{\text{irr}}^2 = \delta\lambda^2\int_0^\tau \dot{g}^2(t)\tau_R[\lambda(t)]\chi[\lambda(t)]dt,
\label{eq:workslow}
\ee
where $\tau_R$ is the relaxation time, defined as
\be
\tau_R = \int_0^\infty \frac{\Psi_0(t)}{\Psi_0(0)}dt,
\label{eq:relaxtime}
\ee
and $\chi$ is the variance of the generalized force, given by
\be
\chi = \beta[\la\partial_\lambda\mathcal{H}^2\ra_0-\la\partial_\lambda\mathcal{H}\ra_0^2],
\label{eq:vargenf}
\ee
or, equivalently,
\be
\chi=\Psi_0(0).
\ee

The parametric dependence on $\lambda(t)$ in the quantities appearing in Eq.~\eqref{eq:workslow} is obtained replacing $\lambda_0$ by $\lambda(t)$ in Eqs. (\ref{eq:relaxtime}) and (\ref{eq:vargenf}). The physical idea behind this is that at each infinitesimal interval of time the system quickly relaxes back to equilibrium such that observables are evaluated instantaneously. Note also that, in contrast to $W_{\text{irr}}^1$, the driving strength $\delta\lambda/\lambda_0$ does not have to be smaller than one.

We want to show that the leading order of $W_{\text{irr}}^2$ in the driving strength $\delta\lambda/\lambda_0$ is equal to $W_{\text{irr}}^1$ (see Eq.~\eqref{eq:wirrfbw}) for large switching times $\tau$. The notation introduced below should be taken as the asymptotic behavior of the corresponding expressions. Thus, the equality
\be
\lim_{\frac{\delta\lambda}{\lambda_0}\ll 1}W_{\text{irr}}^2 = \lim_{\frac{\tau_R}{\tau}\ll 1}W_{\text{irr}}^1
\label{eq:comparison}
\ee
means that the expression of $W^{2}_{\text{irr}}$ for $\delta\lambda/\lambda_{0}\ll 1$ must be equal to the expression of $W^{1}_{\textbf{irr}}$ for $\tau_{R}/\tau\ll 1$. We remark that, in the case of slowly-varying processes, the leading order in $\delta\lambda/\lambda_{0}$ must be the second one, since the integrand, in zeroth order, gives $W^{2}_{\text{irr}}$ proportional to $\delta\lambda^{2}$. This can only happen if there is no dependence of the relaxation time $\tau_{R}$ and variance $\chi$ on the external parameter, that is, if they are calculated at the initial and fixed value $\lambda_{0}$. Thus,
\begin{equation}
\begin{split}
\lim_{\frac{\delta\lambda}{\lambda_0}\ll 1}W_{\text{irr}}^2 &= \delta\lambda^2\int_0^\tau \dot{g}^2(t)\tau_R[\lambda_0]\chi[\lambda_0]dt\\
&= \delta\lambda^2\tau_R[\lambda_0]\chi[\lambda_0]\int_0^\tau \dot{g}^2(t)dt\\
&= \delta\lambda^2\tau_R[\lambda_0]\chi[\lambda_0]\int_0^\tau \int_0^\tau\delta(t-t') \dot{g}(t')\dot{g}(t)dt'dt.\\
\end{split}
\end{equation}

On the other hand, for weak finite-time processes, the irreversible work reads
\be
W_{\text{irr}}^1 = \frac{\delta\lambda^2}{2} \int_0^1\int_0^1 \Psi_0\left(\tau(u-u')\right)\dot{g}(u')\dot{g}(u)du'du.
\label{eq:wirrfbw}
\ee
Now, since the relaxation function vanishes (exponentially for all the cases we have considered here) for large switching times, except when $u=u'$, we have 
\be
\Psi_0\left(\tau(u-u')\right) \sim \xi\delta(\tau(u-u'))
\label{eq:delta}
\ee
for $\tau_{R}/\tau\to 0$. To calculate the factor $\xi$, we use the normalisation of the Dirac delta
\be
\int_{-\infty}^{\infty}\delta(t)dt = 1.
\ee
In this manner, integrating Eq. (\ref{eq:delta}), we have
\be
\xi = 2\tau_R\Psi_0(0) = 2\tau_R[\lambda_0]\chi[\lambda_0],
\ee
which leads to
\be
\lim_{\tau_R/\tau\ll 1}W_{\text{irr}}^{1} = \delta\lambda^2\tau_R[\lambda_0]\chi[\lambda_0]\int_0^\tau \int_0^\tau\delta(t-t') \dot{g}(t')\dot{g}(t)dt' dt,
\ee
proving therefore Eq. (\ref{eq:comparison}). We conclude that the results we obtained for optimal finite-time and weak processes must match the ones obtained in Ref. \cite{deffner2014optimal} in the appropriate limit.


\section{Calculations of $W_{\text{irr}}^{ij}$}
\label{app:C}

The main step to calculate $W_{\text{irr}}^{ij}$  is to perform a change of variables in $t$ and $t'$ such that, in the new variables, the double integral can always be calculated in the domain $[0,\tau_i]\times[0,\tau_j]$. For instance: 
\begin{equation}
\begin{split}
W_{\text{irr}}^{22} = &\, \delta\lambda^2\int_{\tau_1}^{\tau_1+\tau_2}\int_{\tau_1}^t\Psi_0(t-t')\dot{h}_2(t')\dot{h}_2(t)dt'dt\\
= &\, \delta\lambda^2\int_{\tau_1}^{\tau_1+\tau_2}\left[\int_{\tau_1}^t\Psi_0(t-t')\dot{h}_2(t')dt'\right]\dot{g}_2((t-\tau_1)/\tau_2)dt\\
= &\, \delta\lambda^2\int_{0}^{\tau_2}\left[\int_{\tau_1}^{u+\tau_1}\Psi_0(u+\tau_1-t')\dot{h}_2(t')dt'\right]\dot{g}_2(u/\tau_2)du\\
= &\, \delta\lambda^2\int_{0}^{\tau_2}\int_{0}^{u}\Psi_0(u-u')\dot{g}_2(u'/\tau_2)\dot{g}_2(u/\tau_2)du'du.\\
= &\, \frac{\delta\lambda^2}{2}\int_{0}^{\tau_2}\int_{0}^{\tau_2}\Psi_0(u-u')\dot{g}_2(u'/\tau_2)\dot{g}_2(u/\tau_2)du'du.
\end{split}
\label{eq:2prot2}
\end{equation}
Hence, in a similar manner, we have
\be
W_{\text{irr}}^{11} = \frac{\delta\lambda^2}{2}\int_{0}^{\tau_1}\int_{0}^{\tau_1}\Psi_0(u-u')\dot{g}_1(u'/\tau_1)\dot{g}_1(u/\tau_1)du'du,
\label{eq:3prot2}
\ee
\be
W_{\text{irr}}^{21} = \delta\lambda^2 \int_{0}^{\tau_2}\int_{0}^{\tau_1}\Psi_0(u-u'-\tau_1)\dot{g}_1(u'/\tau_1)\dot{g}_2(u/\tau_2)du'du,
\label{eq:3prot2}
\ee
\be
W_{\text{irr}}^{31} = \delta\lambda^2 \int_{0}^{\tau_3}\int_{0}^{\tau_1}\Psi_0(u-u'+\tau_2+\tau_1)\dot{g}_1(u'/\tau_1)\dot{g}_3(u/\tau_3)du'du,
\label{eq:3prot2}
\ee
\be
W_{\text{irr}}^{32} = \delta\lambda^2 \int_{0}^{\tau_3}\int_{0}^{\tau_2}\Psi_0(u-u'+\tau_2)\dot{g}_2(u'/\tau_2)\dot{g}_3(u/\tau_3)du'du,
\label{eq:3prot2}
\ee
\be
W_{\text{irr}}^{33} = \frac{\delta\lambda^2}{2}\int_{0}^{\tau_3}\int_{0}^{\tau_3}\Psi_0(u-u')\dot{g}_3(u'/\tau_3)\dot{g}_3(u/\tau_3)du'du.
\label{eq:3prot2}
\ee

Taking the limits $\tau_1\rightarrow 0$ and $\tau_3\rightarrow 0$, we have
\be
W_{\text{irr}}^{11} = \frac{\delta\lambda^2}{2}\int_{0}^{\tau_1}\int_{0}^{\tau_1}\Psi_0(u-u')\dot{g}_1(u'/\tau_1)\dot{g}_1(u/\tau_1)du'du,
\label{eq:3prot2}
\ee
\be
W_{\text{irr}}^{21} = \delta\lambda^2 g_2(0)\int_0^{\tau_2} \Psi_0(u)\dot{g}_2(u/\tau_2)du,
\label{eq:3prot2}
\ee
\be
W_{\text{irr}}^{22} = \frac{\delta\lambda^2}{2}\int_{0}^{\tau_2}\int_{0}^{\tau_2}\Psi_0(u-u')\dot{g}_2(u'/\tau_2)\dot{g}_2(u/\tau_2)du'du,
\label{eq:3prot2}
\ee
\be
W_{\text{irr}}^{31} = \delta\lambda^2 \Psi_0(\tau)g_2(0)(1-g_2(1)),
\label{eq:3prot2}
\ee
\be
W_{\text{irr}}^{32} = \delta\lambda^2 (1-g_2(1))\int_{0}^{\tau_2}\Psi_0(\tau_2-u)\dot{g}_2(u/\tau_2)du,
\label{eq:3prot2}
\ee
\be
W_{\text{irr}}^{33} = \frac{\delta\lambda^2}{2}\int_{0}^{\tau_3}\int_{0}^{\tau_3}\Psi_0(u-u')\dot{g}_3(u'/\tau_3)\dot{g}_3(u/\tau_3)du'du.
\label{eq:3prot2}
\ee

Expressing all the integrals above in terms of the protocol $g_2(t/\tau_2)$ and summing them, we arrive at Eq. (\ref{eq:wirrtotal}).


\section{Deduction of the backward protocol}
\label{app:D}

In this appendix, we want to express the time-reversed protocol $g_{R}(t)$ in terms of the forward one $g_F(t)$. First, the time-reversed process $\lambda_{R}(t)$ is defined by
\be
\lambda_R(t) = \lambda_F(\tau-t) = \lambda_0+g_F(\tau-t)\delta\lambda.
\label{eq:lambdaback}
\ee
Second, any reversed process that connects the same equilibrium states as $\lambda_{F}(t)$ can be expressed as
\be
\lambda_R(t) = (\lambda_0+\delta\lambda)-g_R(t)\delta\lambda.
\label{eq:lambdaback2}
\ee 
which means that the reversed process drives the system from $\lambda_0+\delta\lambda$ to $\lambda_0$, according to some protocol $g_B(t)$. Expressing Eq. (\ref{eq:lambdaback}) in the following way
\begin{equation}
\begin{split}
\lambda_R(t) & = \lambda_0+g_F(\tau-t)\delta\lambda\\
& = \lambda_0+\delta\lambda-\delta\lambda+g_F(\tau-t)\delta\lambda\\
& = (\lambda_0+\delta\lambda)-[1-g_F(\tau-t)]\delta\lambda,
\end{split}
\end{equation}
we can identify each term with Eq. (\ref{eq:lambdaback2}) and derive Eq. (\ref{eq:backforw}) at the end.

\section{Contributions of the excess power}
\label{app:E}
Using the splitting proposed in Eq.~\eqref{eq:protdiv}, the optimal excess power can be rewritten as
\begin{equation}
\begin{split}
\mathcal{P}_{ex}^*(t,\tau)&=\delta\lambda^2\frac{\dot{g}(t)}{2}\int_0^\tau\Psi_0(t-t')\dot{g}^*(t')dt'\\
&+\delta\lambda^2\frac{g(0^+)}{2}(\Psi_0(\tau-t)+\Psi_0(t))\dot{g}^*(t)\\
&+\delta\lambda^2{g^*}^2(0^+)\Psi(0)(\delta(t)+\delta(\tau-t))
\label{eq:Pextotal}
\end{split}
\end{equation}
The Euler-Lagrange equation \eqref{eq:eleq} can also be rewritten as
\begin{equation}
\begin{split}
\int_0^\tau\Psi_0(t-t')&\dot{g}^*(t')dt'=\int_0^\tau\Psi_0(t')\dot{g}^*(t')dt'\\
&-(\Psi_0(\tau-t)+\Psi_0(t))g^*(0^+)\\
&+(\Psi_0(\tau)+\Psi_0(0))g^*(0^+).
\end{split}
\label{eq:ELmod}
\end{equation}
Substituting Eq.~\eqref{eq:ELmod} in Eq.~\eqref{eq:Pextotal}, we obtain the contributions referring to the continuous and jumps parts.


\bibliographystyle{apsrev4-2}
\bibliography{OPLR.bib}

\end{document}